\documentclass[preprint]{aastex}
\usepackage{color}

\newcommand{\mufun}{$\mu$FUN}

\newcommand{\bdv}[1]{\mbox{\boldmath$#1$}}
\def\bpi{{\bdv{\pi}}}

\author{J.C.~Yee\altaffilmark{1,2}, 
L.-W.~Hung\altaffilmark{3,1,2},
I.A.~Bond\altaffilmark{4,5}, 
W.~Allen\altaffilmark{6,2}, 
L.A.G.~Monard\altaffilmark{7,2}, 
M.D.~Albrow\altaffilmark{8,9},
P.~Fouqu\'e\altaffilmark{10,9},
M.~Dominik\altaffilmark{11,12,13,14}, 
Y.~Tsapras\altaffilmark{15,16,14},
A.~Udalski\altaffilmark{17,18}
% ...
\\ (The MiNDSTEp Consortium)\\
}

\altaffiltext{1}{Department of Astronomy, Ohio State University, 140 West 18th Avenue, Columbus, OH 43210, USA}
\altaffiltext{2}{ Microlensing Follow Up Network ($\mu$FUN) Collaboration}
\altaffiltext{3}{ Department of Physics \& Astronomy, University of California Los Angeles, Los Angeles, CA 90095, USA; liweih@astro.ucla.edu}
\altaffiltext{4}{ Institute for Information and Mathematical Sciences, Massey University, Private Bag 102-904, Auckland 1330, New Zealand}
\altaffiltext{5}{ Microlensing Observations in Astrophysics (MOA) Collaboration}
\altaffiltext{6}{ Vintage Lane Observatory, Blenheim, New Zealand}
\altaffiltext{7}{ Bronberg Observatory, Centre for Backyard Astrophysics, Pretoria, South Africa}
\altaffiltext{8}{University of Canterbury, Department of Physics and Astronomy, Private Bag 4800, Christchurch 8020, New Zealand}
\altaffiltext{9}{Probing Lensing Anomalies Network (PLANET) Collaboration}
\altaffiltext{10}{ IRAP, CNRS, Universit\'{e} de Toulouse, 14 avenue Edouard Belin, 31400 Toulouse, France}
\altaffiltext{11}{SUPA, University of St Andrews, School of Physics \& Astronomy, North Haugh, St Andrews, KY16 9SS, UK}
\altaffiltext{12}{Royal Society University Research Fellow}
\altaffiltext{13}{Microlensing Network for the Detection of Small Terrestrial Exoplanets (MiNDSTEp) Consortium}
\altaffiltext{14}{RoboNet Collaboration}
\altaffiltext{15}{Las Cumbres Observatory Global Telescope Network, 6740B Cortona Dr, Goleta, CA 93117, USA}
\altaffiltext{16}{School of Physics and Astronomy, Queen Mary University of London, Mile End Road, London, E1 4NS}
\altaffiltext{17}{Warsaw University Observatory, Al. Ujazdowskie 4, 00-478 Warszawa, Poland}
\altaffiltext{18}{Optical Gravitational Lensing Experiment (OGLE)}
% ...

\title{MOA-2010-BLG-311: A planetary candidate below the threshold of
reliable detection}

\begin{document}
\begin{abstract}
  We analyze MOA-2010-BLG-311, a high magnification ($A_{\rm
    max}>600$) microlensing event with complete data coverage over the
  peak, making it very sensitive to planetary signals. We fit this
  event with both a point lens and a 2-body lens model and find that
  the 2-body lens model is a better fit but with only
  $\Delta\chi^2\sim 80$. The preferred mass ratio between the lens star and its
  companion is $q=10^{-3.7\pm0.1}$, placing the candidate companion in
  the planetary regime. Despite the formal significance of the planet,
  we show that because of systematics in the data the evidence for a
  planetary companion to the lens is too tenuous to claim a secure
  detection. When combined with analyses of other high-magnification
  events, this event helps empirically define the threshold for
  reliable planet detection in high-magnification events, which
  remains an open question.
\end{abstract}

\keywords{Galaxy: bulge --- gravitational lensing: micro --- planetary
systems: detection}

\clearpage
{\section{Introduction}
\label{sec:intro}}

High-magnification events, events in which the maximum magnification
of the source, $A_{\rm max}$, is greater than 100, have been a major
focus of microlensing observations and analysis. Because the impact
parameter between the source and the lens is very small in such cases,
$u_0\simeq 1/A_{\rm max}$, it is likely to probe a central caustic
produced by a planetary companion to the lens star. Furthermore, such
events can often be predicted in advance of the peak, allowing
intensive observations of the event at the time when it is most
sensitive to planets. Consequently, a substantial amount of effort has
been put into identifying, observing, and analyzing such events.

Observed high-magnification events are classified into two groups for
further analysis: events with signals obvious to the eye and events
without. Only events in the first category are systematically fit with
2(or more)-body models. The other events are only analyzed to
determine their detection efficiencies. As a result, no planets have
been found at or close to the detection threshold, and furthermore
this detection threshold is not well understood\footnote{The need for
a well-defined detection threshold is also discussed in
\citealt{Yee12}.}. \citet{Gould10} suggest a detection threshold in
the range of $\Delta\chi^2=\,$350--700 is required to both detect the
signal and constrain it to be planetary, but they note that the exact
value is unknown. With the advent of second-generation microlensing
surveys, which will be able to detect planets as part of a controlled
experiment with a fixed observing cadence, it is important to study
the reliability of signals close to the detection threshold, since a
systematic analysis of all events in such a survey will yield signals
of all magnitudes, some of which will be real and some of which will
be spurious.

In this paper, we present the analysis of a high-magnification
microlensing event, MOA-2010-BLG-311, which has a planetary signal
slightly too small to claim as a detection. We summarize the data
properties in Section \ref{sec:data} and present the color-magnitude
diagram (CMD) in Section \ref{sec:cmd}. In Section \ref{sec:models},
we fit the light curve with both point lens and 2-body models. We then
discuss why a planetary detection cannot be claimed in Section
\ref{sec:reliability}. We calculate the Einstein ring size and
relative proper motion in Section \ref{sec:temu} and discuss the
possibility that the lens is a member of the cluster NGC 6553 in
Section \ref{sec:cluster}. We give our conclusions in Section
\ref{sec:discuss}.

{\section{Data}
\label{sec:data}}
\subsection{Observations}

On 2010 June 15 (HJD$'$ 5362.967 $\equiv$ HJD$-2450000$), the
Microlensing Observations in Astrophysics (MOA) collaboration
\citep{Bond01, Sumi11} detected a new microlensing event
MOA-2010-BLG-310 at (R.A., decl.) = (18$^{\rm h}$08$^{\rm
m}$49.\hskip-2pt$^{\rm s}$98, -25$^{\circ}57'04.\hskip-2pt''27$)
(J2000.0), (l, b) = (5.17, -2.96), along our line of sight toward the
Galactic Bulge. MOA announced the event through its email alert system
and made the data available in real-time. Within a day, this event was
identified as likely to reach high magnification. Because of MOA's
real-time alert system, the event was identified sufficiently far in
advance to allow intensive follow up observations over the peak.

The observational data were acquired from multiple observatories,
including members of the MOA, OGLE, \mufun , PLANET, RoboNet
\citep{Tsapras09}, and MiNDSTEp collaborations. In total, sixteen
observatories monitored the event for more than one night, and thus
their data were used in the following analysis. Among these, there is
the MOA survey telescope (1.8m, MOA-Red\footnote{This custom filter
  has a similar spectral response to $R$-band}, New Zealand) and the
B\&C telescope (60cm, $V$, $I$, New Zealand); eight of the
observatories are from \mufun: Auckland (AO, 0.4m, Wratten \#12, New
Zealand), Bronberg (0.36m, unfiltered, South Africa), CTIO SMARTS
(1.3m, $V$, $I$, $H$, Chile), Farm Cove (FCO, 0.36m, unfiltered, New
Zealand), Kumeu (0.36m, unfiltered, New Zealand), Molehill (MAO, 0.3m,
unfiltered, New Zealand), Vintage Lane (VLO, 0.4m, unfiltered, New
Zealand), and Wise (0.46m, unfiltered, Israel); three are from PLANET:
Kuiper telescope on Mt. Bigelow (1.55m, $I$, Arizona), Canopus (1.0m,
$I$, Australia), and SAAO (1.0m, $I$, South Africa); one is from
RoboNet: Liverpool (2.0m, $I$, Canaries); and one is from MiNDSTEp: La
Silla (1.5m, $I$, Chile). The event also fell in the footprint of the
OGLE IV survey (1.3m, $I$, Chile), which was in the commissioning
phase in 2010. The observatory and filter information is summarized in
Table \ref{tab:data}.

In particular, observations from the MOA survey telescope, MOA B\&C,
PLANET Canopus, \mufun\ Bronberg, and \mufun\ VLO provided nearly
complete coverage over the event peak between HJD$' = 5365.0$ and
HJD$' = 5365.4$.

\subsection{Data Reduction}

The MOA and B\&C data were reduced with the standard MOA pipeline
\citep{Bond01}. The data from the \mufun\ observatories were reduced
using the standard DoPhot reduction \citep{Schechter93}, with the
exception of Bronberg and VLO data, which were reduced using
difference image analysis \citep[DIA;][]{Alard00,Wozniak00}.  Data
from the PLANET and RoboNet collaborations were reduced using pySIS2
pipeline \citep{Bramich08,Albrow09}. Data from MiNDSTEp were also
initially reduced using the pySIS2 pipeline.  The OGLE data were
reduced using the standard OGLE pipeline \citep{Udalski03}. Both the
MOA and MiNDSTEp/La Silla data were reduced in real-time, and as such
the initial reductions were sub-optimal. In fact, the original MOA
data over the peak were unusable because they were corrupted. After
the initial analysis, both the MiNDSTEp and MOA data were rereduced
using optimized parameters.

{\section{Color-Magnitude Diagram}
\label{sec:cmd}}

\begin{figure}
\includegraphics{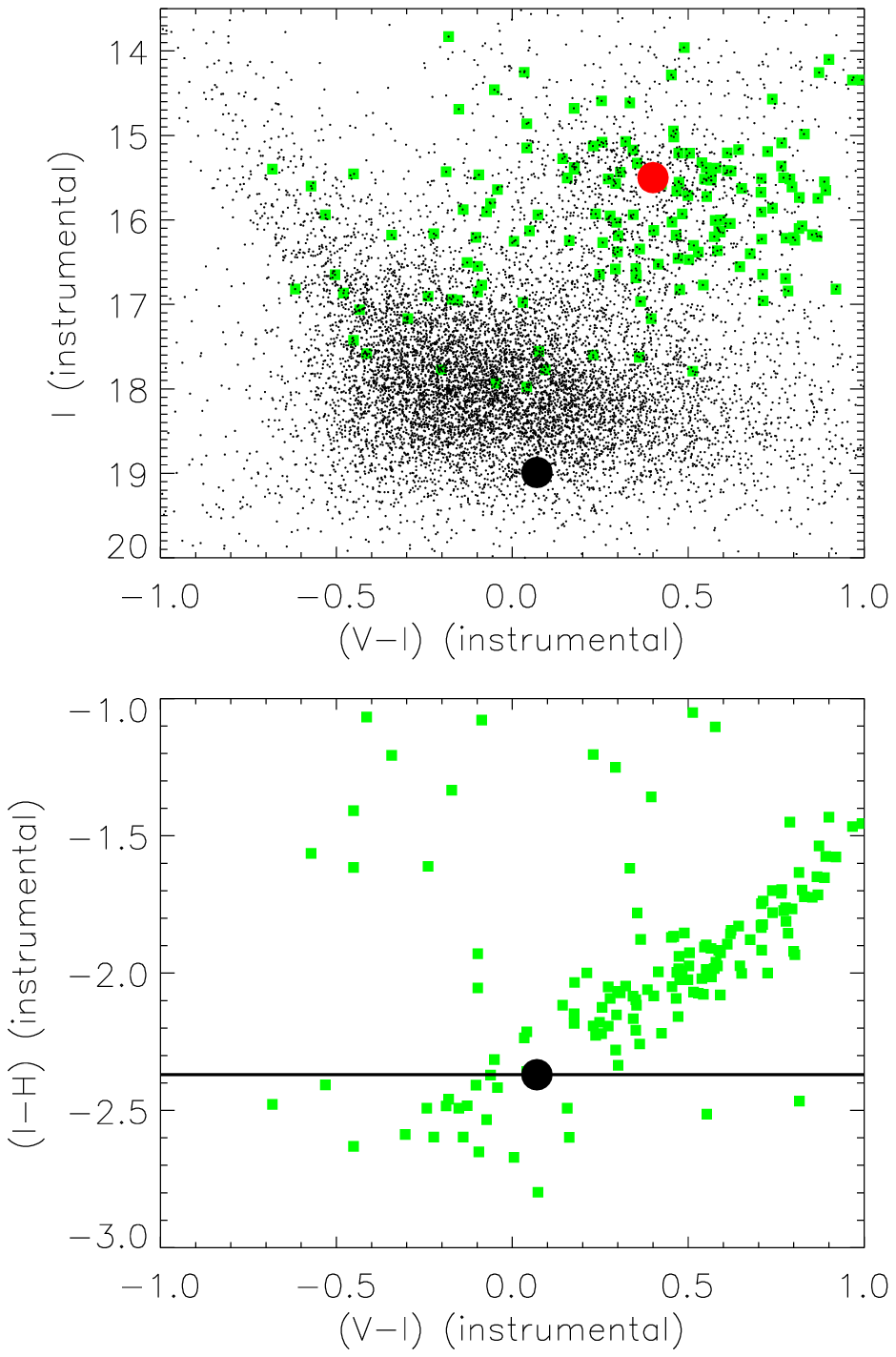}
\caption{Color-magnitude diagram (top) and color-color diagram
(bottom) of the field of view of MOA-2010-BLG-311 constructed from the
CTIO observations. The small black dots represent the field stars;
stars used in constructing the color-color diagram are shown as the
green squares in both panels. The black line in the bottom plot
shows the measured instrumental $(I-H)$ color of the
source. The $(V-I)$ color of the source is then derived from the
intersection of this line with the $(V-I)$---$(I-H)$ relation
established by the field stars (green squares), producing the black
dot in the bottom panel. The position of the source in the CMD (large
black dot) is given by this $(V-I)$ measurement and the source flux
from the fits to the light curve. The large red dot shows the centroid of
the red clump. \label{fig:cmd}}
\end{figure}

To determine the intrinsic source color, we construct a
color-magnitude diagram (CMD) of the field of view containing the
lensing event (Figure~\ref{fig:cmd}) based on $V$- and $I$-band images
from CTIO SMARTS ANDICAM camera. The field stars in the CMD are
determined from three $V$-band images and multiple $I$-band
images. Four $V$-band images were taken; however, only three of the
images are of sufficient quality to contribute to the CMD. We visually
checked each of the three images to make sure that there were no
obvious defects such as cosmic-ray events in the images. 

From the fit to the light curve, we find the instrumental magnitude of the
source is $I=19.0$ with an uncertainty of 0.05 mag due to
differences between the planet and point lens models as well as the
error parameterizations. In the top right-hand corner of
Figure~\ref{fig:cmd}, the red dot at $(V-I, I)_{\rm cl}=(0.40,
15.5)$ marks the centroid of the red clump. The intrinsic color and
magnitude of the clump are $(V-I, I)_{\rm cl,0}=(1.06,14.3)$
\citep{Bensby11, Nataf12}. Using the offset between the intrinsic
magnitude of the clump and the observed, instrumental magnitude, we
can then calibrate the magnitude of the source to find
$I_{S,0}=17.8\pm0.1$.

The color of the source is normally estimated from $V$- and $I$-band
images using the standard technique in \citet{Yoo04}. However, with
only one highly-magnified $V$-band image, this method is unreliable,
so we use an alternative technique to determine the instrumental
$(V-I)$ color by converting from the instrumental $(I-H)$ color. Using
the simultaneous CTIO $I$- and $H$-band observations, we measure the
instrumental $(I-H)$ color of the source by linear regression of $H$
on $I$ flux at various magnifications during the event.  We then
construct a $VIH$ instrumental color-color diagram from stars in the
field (bottom panel of Figure \ref{fig:cmd}). The stars are chosen to
be all stars seen in all three bands with instrumental magnitude
brighter than $H_{\rm CTIO}=19.0$ (note that the field of view for $H$-band
observations is $2.4'\times2.4'$ compared to $6'\times6'$ for the
optical bands). The field stars form a well defined track, which
enables us to estimate the $(V-I)$ source color from the observed
$(I-H)$ source color.  This yields $(V-I)_0=0.75\pm0.05$.  Note that
this method would not work for red stars, $(V-I)_0>1.3$, because for
these red stars, the $VIH$ relation differs between giants and dwarfs
\citep{Bessell88}. However, the observed color is well blueward of
this bifurcation. There is also a spectrum of the source taken at
HJD$^{\prime}=5365.001$ \citep{Bensby11}. The ``spectroscopic''
$(V-I)_0$ reported in that work is 0.77, in good agreement with the
value calculated here.

{\section{Modeling}
\label{sec:models}}

{\subsection{The Basic Model}}

A casual inspection of the light curve does not show any deviations
from a point lens, so we begin by fitting a point lens model to the
data. A point lens model is characterized by three basic parameters:
the time of the peak, $t_0$, the impact parameter between the source
and the lens stars, $u_0$, and the Einstein timescale, $t_E$. Since
$u_0$ is small, the finite size of the source can be important. To
include this effect, we introduce the source size in Einstein radii,
$\rho$, as a parameter in the model. Additionally, we include
limb-darkening of the source. The temperature of this slightly evolved
source was determined from the spectrum to be $T_{\rm eff} \sim$ 5460
K by \citet{Bensby11}. Using \citet{Claret00} we found the
limb-darkening coefficients to be $u_V = 0.7086$, $u_I = 0.5470$, and
$u_H = 0.3624$, assuming a microturbulent velocity = 1 km s$^{-1}$,
log $g$ = 4.0, solar metallicity, and $T_{\rm eff} = 5500$K, which is
the closest grid point given the \citet{Bensby11} measurements. We
then convert $u_V$, $u_I$, and $u_H$ to the form introduced by
\citet{Albrow99}
\begin{equation}
\label{eqn:limbdcoeff}
\Gamma = \frac{2u}{3-u}
\end{equation}
to obtain $\Gamma _V = 0.62$, $\Gamma _I = 0.45$, and $\Gamma _H =
0.28$. Because the various data sets are not on a common flux scale,
there are also two flux parameters for each data set, $f_{\rm S,i}$ and
$f_{\rm B,i}$, such that
\begin{equation}
f_{{\rm mod},i}= f_{{\rm S},i}A(t)+f_{{\rm B},i},
\end{equation}
where $A(t)$ is the predicted magnification of the model at time, $t$,
and includes the appropriate limb-darkening for data set $i$. The
source flux is given by $f_{{\rm S}, i}$, and $f_{{\rm B}, i}$ is the
flux of all other stars, including any light from the lens, blended
into the PSF (i.e., the ``blend'').


{\subsection{Error Renormalization}
\label{sec:errors}}

As is frequently the case for microlensing data, the initial point
lens fit reveals that the errors calculated for each data point by the
photometry packages underestimate the true errors. Additionally, there
are be outliers in the data that are clearly seen to be spurious by
comparison to other data taken simultaneously from a different
site. Simply taking the error bars at face value would lead to biases
in the modeling. Because the level of systematics varies between
different data sets, underestimated error bars can give undue weight
to a particular set of data. Additionally, if the errors are
underestimated, the relative $\Delta\chi^2$ between two models will be
overestimated, making the constraints seem stronger than they actually
are.

To resolve these issues, we rescale the error bars using an error
renormalization factor (or factors) and eliminate outliers. We begin
by fitting the data to a point lens to find the error renormalization
factors. We remove the outliers according to the procedure described
below based on the renormalized errors, refit, and recalculate the
error renormalization factors. We repeat this process until no further
outliers are found. 

To first order, we can compensate for the underestimated error bars by
rescaling them by a single factor, $k$. The rescaling factor is chosen
for each data set, $i$, such that $k_i=(\chi_i^2/N_i)^{1/2}$,
producing a $\chi^2$ per degree of freedom of $\chi^2_{\rm dof}=1$. We
will refer to this simple scheme for renormalizing the errors as
``1-parameter errors''.

Alternatively, we can use a more complex method to renormalize the
errors, which we will call ``2-parameter errors''. This method was
also used in \citet{Miyake12} and \citet{Bachelet12}. To renormalize
the errors, we rank order the data by magnification and calculate two
factors, $k$ and $e_{\rm min}$, such that
\begin{equation}
\sigma_{j}=k_i\sqrt{\sigma_{{\rm orig}, j}^2+e_{{\rm min}, i}^2},
\end{equation}
where $\sigma_{\rm orig}$ is the original error bar and $\sigma$ is
the new error bar and the calculations are done in magnitudes. The
index $i$ refers to a particular data set, and $j$ refers to a
particular point within that data set. The additional term, $e_{\rm
min}$, enforces a minimum uncertainty in magnitudes, because at high
magnification, the flux is large, so the formal errors on the
measurement can be unrealistically small. The error factors, $k$ and
$e_{\rm min}$, are chosen so that $\chi^2_{\rm dof}$ for points sorted
by magnification increases in a uniform, linear fashion and
$\chi^2_{\rm dof}=1$ \citep{Yee12}. Not all data sets will require an
$e_{\rm min}$ term; it is only necessary in cases for which
$\chi^2_{\rm dof}$ has a break because the formal errors are too small
when the event is bright. Note that the $e_{\rm min}$ factor will be
primarily affected by the points taken when the event is bright,
whereas $k$ is affected by all points, so if there are many more
points at the baseline of the event, these will dominate the
calculation of $k$.

To remove outliers in the data we begin by eliminating any
points taken at airmass $>3$ or during twilight.
Additional outliers are defined as any point more than $X\sigma$ from
the expected value, where $X$ is determined by the number of data
points, such that fewer than one point is expected to be more than
$X\sigma$ from the expected value assuming a Gaussian error
distribution. The normal procedure is to compare the data to the
``expected value'' from a point lens fit. We do this for data in the
wings of the event, $t($HJD$^{\prime})<5363$ or
$t($HJD$^{\prime})>5367$, when we do not expect to see any real signals.
However, the peak of the event, $5365 \leq t($HJD$^{\prime})\leq 5367$, is
when we would expect to see a signal from a planet if one exists. A
planetary signal would necessarily deviate from the expectation for a
point lens, so in this region instead of comparing to a point lens
model, we use the following procedure to identify outliers:
\begin{enumerate}

\item{For each point, we determine whether there are points from any
other data set within 0.01 days. If there are more than two points
from a given comparison dataset in this range, we keep only the
point(s) immediately before and/or after the time of the point in
question (i.e., a maximum of 2 points). The point is not compared to other
points from the same data set.\label{step3}}

\item{If there are matches to at least two other data sets, we proceed
to (\ref{step6}) to determine whether or not the point is an
outlier. Otherwise, we treat the point as good.}

\item{We then determine the mean of the collected points, $\mu$, including
the point in question, by maximizing a likelihood function for data with
outliers \citep{Sivia10}:
\begin{equation}
L \propto \sum_{j=1}^N \ln\left(\frac{1-e^{-R_j^2/2}}{R_j^2}\right)
\end{equation}
where $R_j=(\mu-x_j)/\sigma_j$ and $x_j$ is the datum and $\sigma_j$
is its renormalized error bar. If the flux is changing too rapidly for
the points to be described well by a mean, we fit a line to the data.
\label{step6}}

\item{We compare the point to the predicted value, using the
likelihood function to see if it is an outlier:

   (a) If there is only one likelihood maximum, we calculate $R_j$. If
   $R_j>X$, the point is rejected as an outlier.

   (b) If there is more than one maximum, and the point in question
   falls in the range spanned by the maxima, we assume the point is
   good. If it falls outside the range, we calculate $R_j$ using the
   nearest maximum to determine $\mu$. If $R_j>X$, we flag the point
   as an outlier.}

\end{enumerate}
Note that for this procedure, we use the point lens model values of
$F_{\rm S,i}$ and $F_{\rm B, i}$ to place all of the data sets on a
common flux scale. 

This procedure is more complicated than usual, but because we compare
the points only to other data, rather than some unknown model, it
provides an objective means to determine whether a point is an outlier
without destroying real signals corroborated by other data. We also
visually inspect each set of points to confirm that the algorithm
works as expected. Because finite source effects are significant in
this event, one might expect slight differences among data sets due to
the different filters, so as part of this visual inspection we also
checked that this did not play a significant role.

For both 1-parameter and 2-parameter errors, Table \ref{tab:data}
lists the error normalization factors and number of points for each
observatory that survived the rejection process. Which of these error
parameterizations correctly describes the data depends on the nature
of the underlying errors. In principle, the noise properties of the
data are fully described by a covariance matrix of all data points,
but we are unable to calculate such a matrix. Instead, we have two
different error parameterizations. The $k$ factor is calculated
primarily based on baseline data for which statistical errors
dominate. In contrast, the $e_{\rm min}$ factor is heavily influenced
by points at the peak of the event when systematic errors are
important. Thus, 2-parameter errors better reflect the systematic
errors, whereas 1-parameter errors better reflect the statistical
errors.

Correlated errors often have a major impact on our ability to
determine whether or not the planetary signal in this event is
real. We know that correlations in the microlensing data exist, but
there has not been a systematic investigation of this in the
microlensing literature. Correlated errors (red noise) are generally
thought of as reducing the sensitivity to signals, because successive
points are not independent, giving related information. But in fact,
sharp, short timescale signals are not degraded by correlated noise
and may still be robustly detected.

Consider the case of a short-timescale signal superimposed on a
long-timescale correlation. Then a model may reproduce the
short-timescale signal leading to an improvement in $\chi^2$ without
actually passing through the data because of the overall offset caused
by the correlations. Now suppose that the correlated, red noise has a
larger-amplitude than the white noise (i.e., statistically
uncorrelated noise). If we set the error bars by the large-amplitude
deviations, which is correct for long timescales for which the data
are uncorrelated, the significance of the short timescale jump will be
diluted, possibly to the point of being considered statistically
insignificant. However, if the timescale of the signal is much shorter
than the correlation length of the red noise, the significance of the
signal should actually be judged against the white noise, since on
that timescale, the red noise will only contribute a constant offset.

In this case, we expect the planetary signal to be quite short, so if
the systematic noise is dominated by correlated errors, the noise
should be better described by the 1-parameter errors. Because the
source in this event crosses the position of the lens and there are no
obvious deviations due to a planet, we expect that any planetary
signals will be due to very small caustics, which are detectable only
at the limb-crossing times ($t_{\rm limb}=t_0\pm
t_E\sqrt{\rho^2-u_0^2}$) when the caustic passes onto and off-of the
face of the source. Therefore, the timescale of such a perturbation
will be very short, equal to $t_{\rm E}$ times the caustic size $w$,
which is $\lesssim 15$ minutes. In contrast, observed correlations in
the microlensing data are typically on longer timescales,
$\mathcal{O}($hour$)$ (based on our experience with microlensing data
which are usually sampled with a frequency of $\sim 15$
minutes). Hence, the timescale of the signal is likely to be less than
the timescale of the correlated noise.

However, there are other sources of systematic errors that are
unrelated to correlated noise such as flat-fielding errors. If such
errors dominate over correlated noise, then the 2-parameter errors are
a better description of the error bars over the peak.

Because the systematic errors, correlated and uncorrelated, have not
been studied in detail, we are unable to determine which is the
dominant effect. Hence, we are also unable to determine which error
prescription better describes our data. We will begin by analyzing the
light curve using 1-parameter errors. We will then discuss how the
situation changes for 2-parameter errors.

{\subsection{Point Lens Models}
\label{sec:pl}}

\begin{figure}
\includegraphics[width=\textwidth]{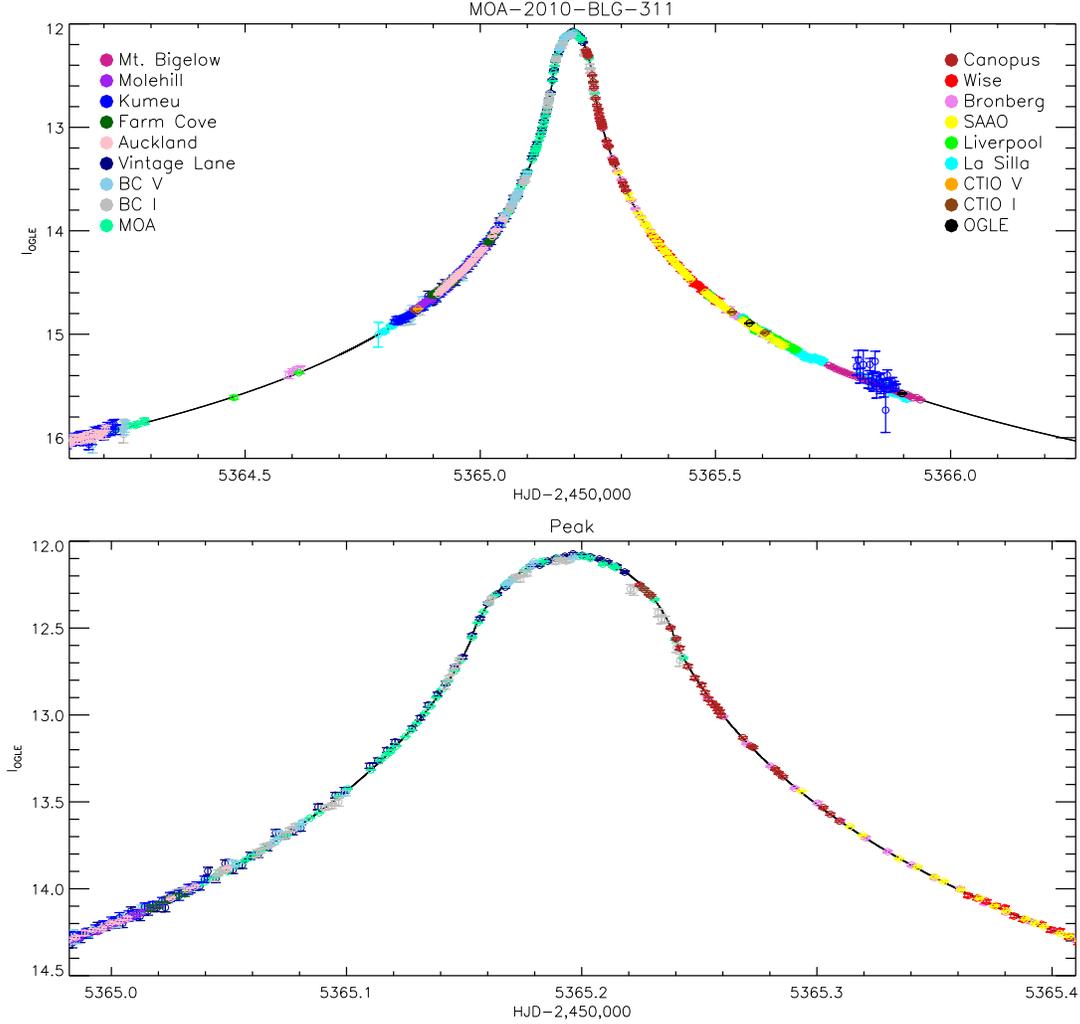}
\caption{Light curve of MOA-2010-BLG-311. Data from different
observatories are plotted in different colors. The data from Bronberg
({\it medium pink}) have been binned for clarity in the figures; only
unbinned data were used in the fitting. The black line shows the
best-fit point lens model; on this scale, the best-fit planetary model
appears very similar. The error bars reflect 1-parameter errors
(see Sec. \ref{sec:errors}).\label{fig:lc} }
\end{figure}

\begin{figure}
\includegraphics[width=\textwidth]{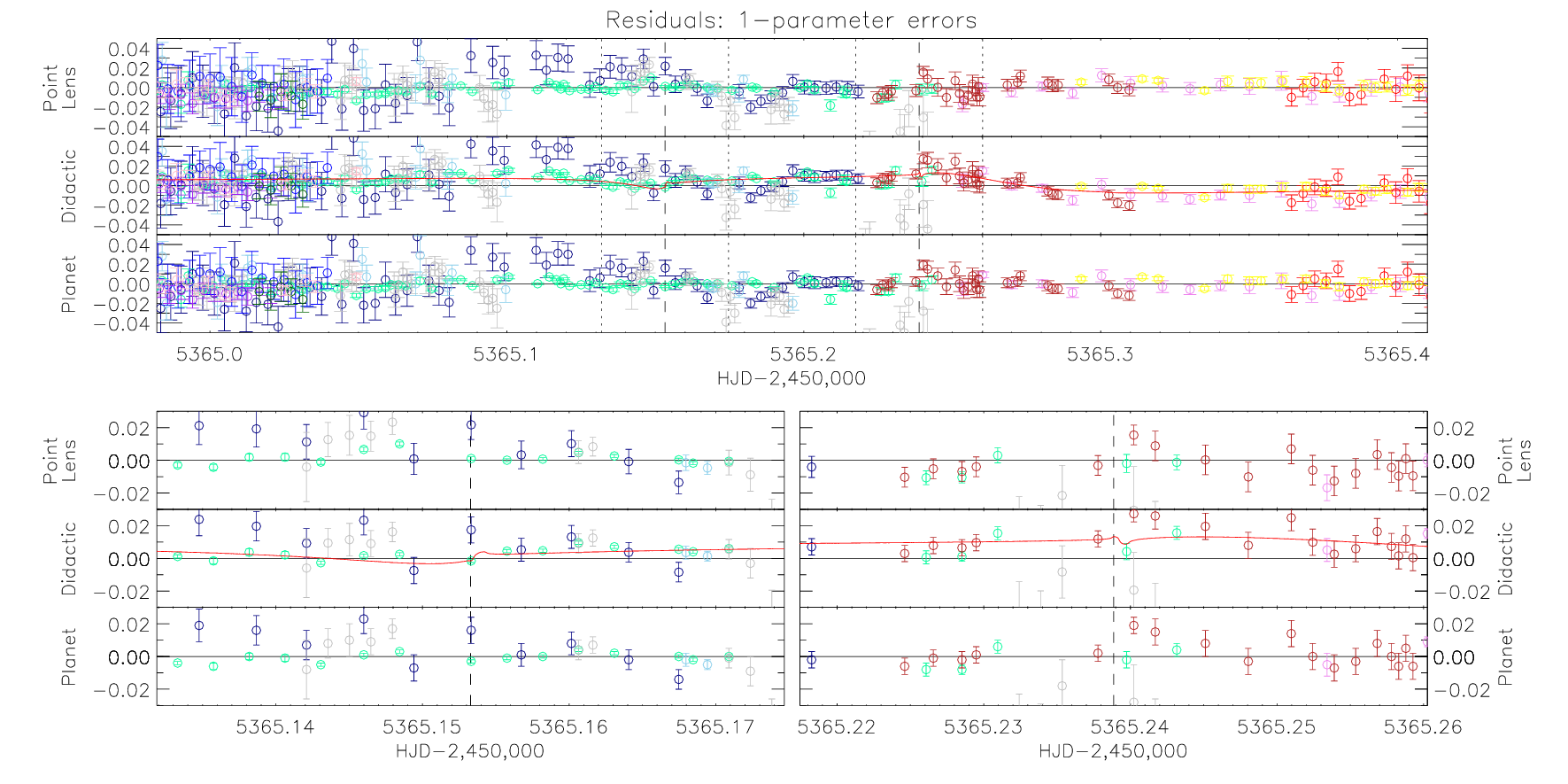}
\includegraphics[width=\textwidth]{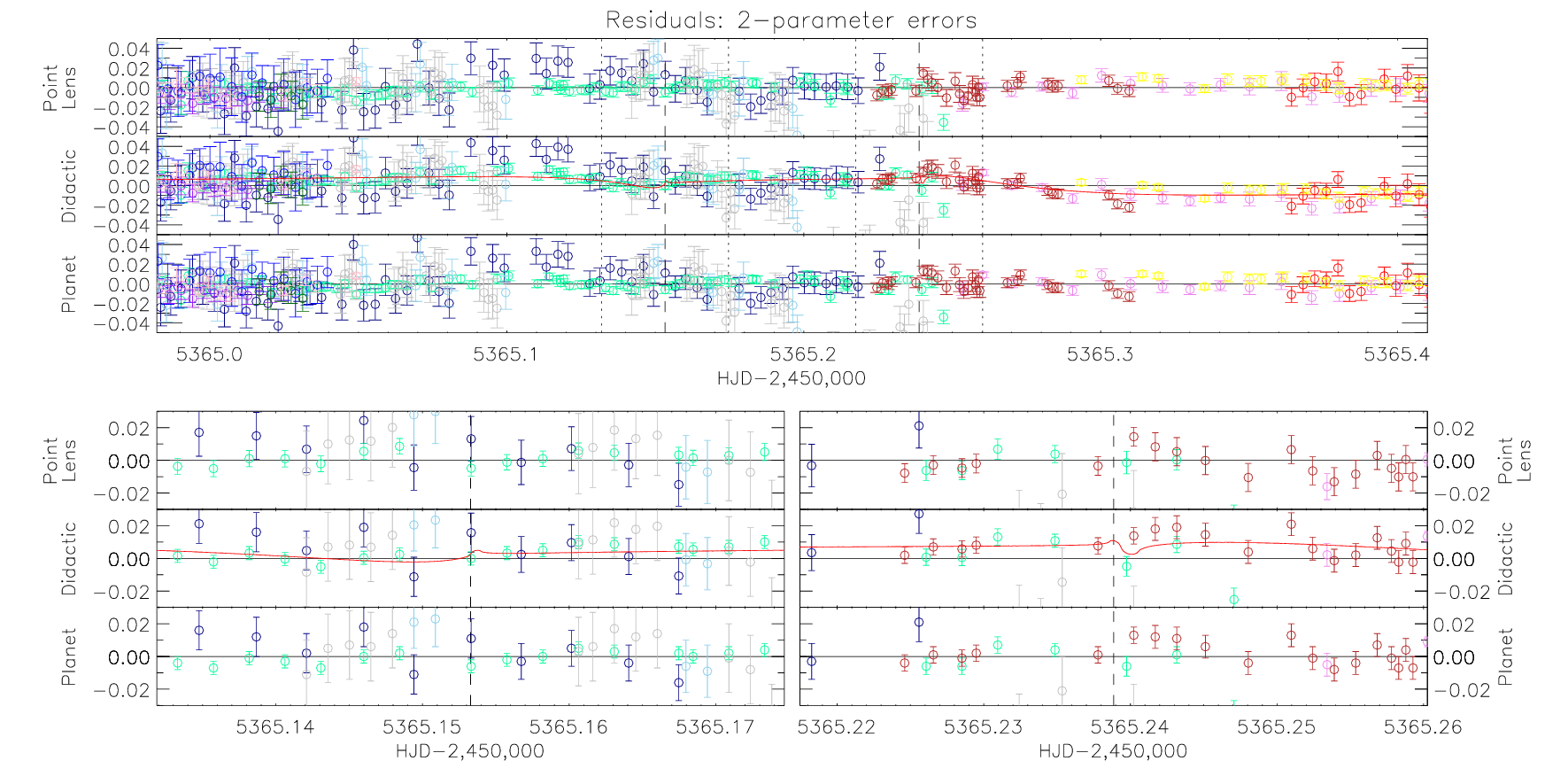}
\caption{Residuals over the peak to the best-fit point lens and
planetary microlensing models. The top set of panels shows 1-parameter
errors, and the bottom set shows 2-parameter errors. Note that for the
time range shown, which is the same as for the bottom panel for
Fig. \ref{fig:lc}, the error bars tend to be larger for 2-parameter
errors than for 1-parameter errors. Careful inspection of the
residuals to the point lens and planet models shows improvement around
the times of the limb-crossings of the source star and also on the
falling side of the light curve. In the middle panel of each set, the
red lines show the difference between the best planetary and best
point lens models. The didactic residuals in the middle panel are the
sum of the red line and the residuals to the best planet model (bottom
panel). The dashed lines indicate the limb-crossing times. The dotted
lines in the top panels of each set indicate the time ranges shown in the
bottom panels. Note that the scales for the top and bottom residuals
panels are not the same. The colors of the points are the same as in
Fig. \ref{fig:lc}. The Bronberg data have
been binned for clarity (as in Fig. \ref{fig:lc})\label{fig:res}}.
\end{figure}

The best-fit point lens model and the uncertainties in the parameters
are given in Table \ref{tab:models1}. This model is shown in Figure
\ref{fig:lc}, and the residuals to this fit are shown in Figure
\ref{fig:res}. These exhibit no obvious deviations. These fits confirm
that finite source effects are important, since $\rho$ is well
measured and larger than the impact parameter, $u_0$. 

We also fit a point lens model that includes the microlens parallax effect,
which arises either from the orbital motion of the Earth during the
event or from the difference in sightlines from two or more
observatories separated on the surface of the Earth. Microlens
parallax enters as a vector quantity: $\bpi_{\rm E}=(\pi_{\rm E,N},
\pi_{\rm E, E})$. The addition of parallax can break the degeneracy
between solutions with $u_0>0$ and $u_0<0$, so we fit both
cases. Parallax does improve the fit beyond what is expected simply
from adding 2 more free parameters and shows a preference for
$u_0>0$. However, we shall see in the next section that a planetary
model without parallax produces an even better fit and adding
parallax in addition to the planet gives only a small additional
improvement.

{\subsection{2-body Models}
\label{sec:planet}}

We search for 2-body models over a broad range of mass ratios, from
$q=10^{-6}$ to $q=10^{-1}$. For each value of $q$, we chose a range
for the projected separation between the two bodies, $s$, for which
the resulting caustic is smaller than $\rho$ and $s<1.0$. For each
combination of $s$ and $q$, we allow the angle of the source
trajectory, $\alpha$, to vary, seeding each run with values of
$\alpha$ from 0 to 360 degrees in steps of 5 degrees. For our models,
we use the map-making method of \citet{Dong06} when the source is
within two source radii of the position of the center of
magnification. Outside this time range, we use the hexadecapole or
quadrupole approximations for the magnification
\citep{Pejcha09,Gould08}. We used a Markov Chain Monte Carlo to find
the best-fit parameters and uncertainties for each $s$, $q$
combination.

\begin{figure}
\includegraphics{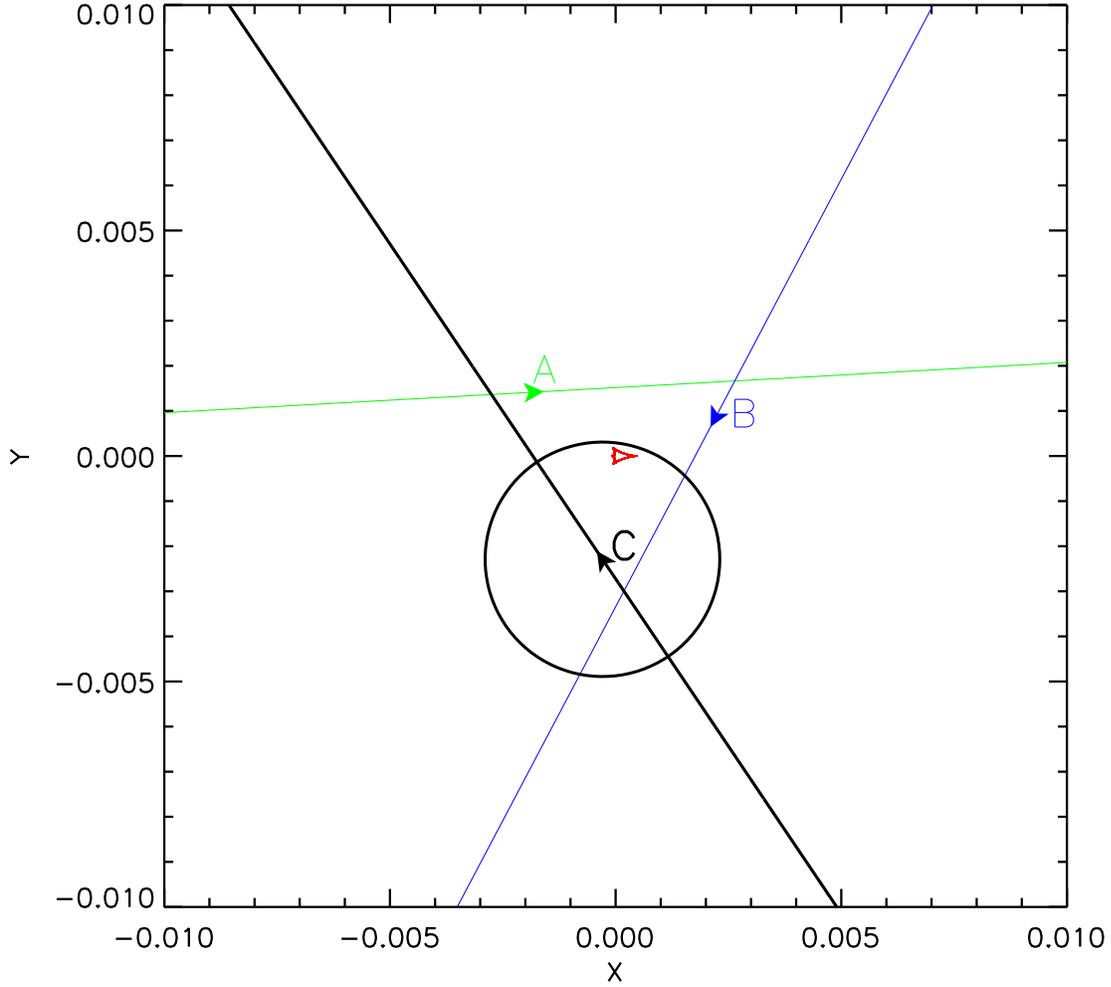}
\caption{The caustic (red) and three source trajectories corresponding
to the three minima discussed in Section \ref{sec:planet}. Trajectory
C (black) is preferred by $\Delta\chi^2\gtrsim35$ over trajectories A
(green) and B (blue) for models with 1-parameter errors and without
parallax. The source is shown to scale as the large circle; the lines
with arrows indicate the trajectories of the center of the source. The
abscissa in these plots is parallel to the binary axis with the lens
star close to the origin and planet to the right. The scale is such
that 1.0 equals the Einstein radius.\label{fig:traj}}
\end{figure}

The grid search reveals an overall improvement in $\chi^2$ relative to
the point lens model. We find three $\chi^2$ minima for different
angles for the source trajectory. For central caustics with planetary
mass ratios, the caustic is roughly triangular in shape with a fourth
cusp where the short side of the triangle intersects the binary axis;
the three trajectories roughly correspond to the three major cusps of
the caustic. An example caustic is shown in Figure \ref{fig:traj}
along with the trajectories corresponding to the three minima. The
angles of the three trajectories are approximately $\alpha=0, 115,$ and
235 degrees, and we will refer to them as trajectories ``A'', ``B'',
and ``C'', respectively.

We then refine our grid of $s$ and $q$ around each of these three
minima. We repeat these fits accounting for various microlensing
degeneracies. First, we fit without parallax and assuming $s<1$. Then,
we add parallax and fit both with $u_0>0$ and $u_0<0$ to see if this
degeneracy is broken. Finally, we fit 2-body lens models with $s>1$
and no parallax, since there is a well known microlensing degeneracy
that takes $s\rightarrow s^{-1}$. 

The best-fit solution has $\chi^2=6637.96$ and $\alpha= 236.4$. This
reflects an improvement in $\chi^2$ of $\Delta\chi^2\sim140$ over the
point lens solution. There is no preference for $s<1$ over $s>1$, but
trajectory C is preferred by $\Delta\chi^2\gtrsim35$ over trajectories
A and B. The parameters and their uncertainties for the planet fits are given
in Table \ref{tab:models1}. The mass ratio between the lens star and
its companion is firmly in the planetary regime:
$q=10^{-3.7\pm0.1}$. Furthermore, planetary mass ratios are clearly
preferred over ``stellar'' mass ratios ($q\sim0.1$), which are
disfavored by more than $\Delta\chi^2=60$. Parallax further improves the
fit by only $\Delta\chi^2\sim10$ and has little effect on the other
parameters.

\begin{figure}
\includegraphics[width=\textwidth]{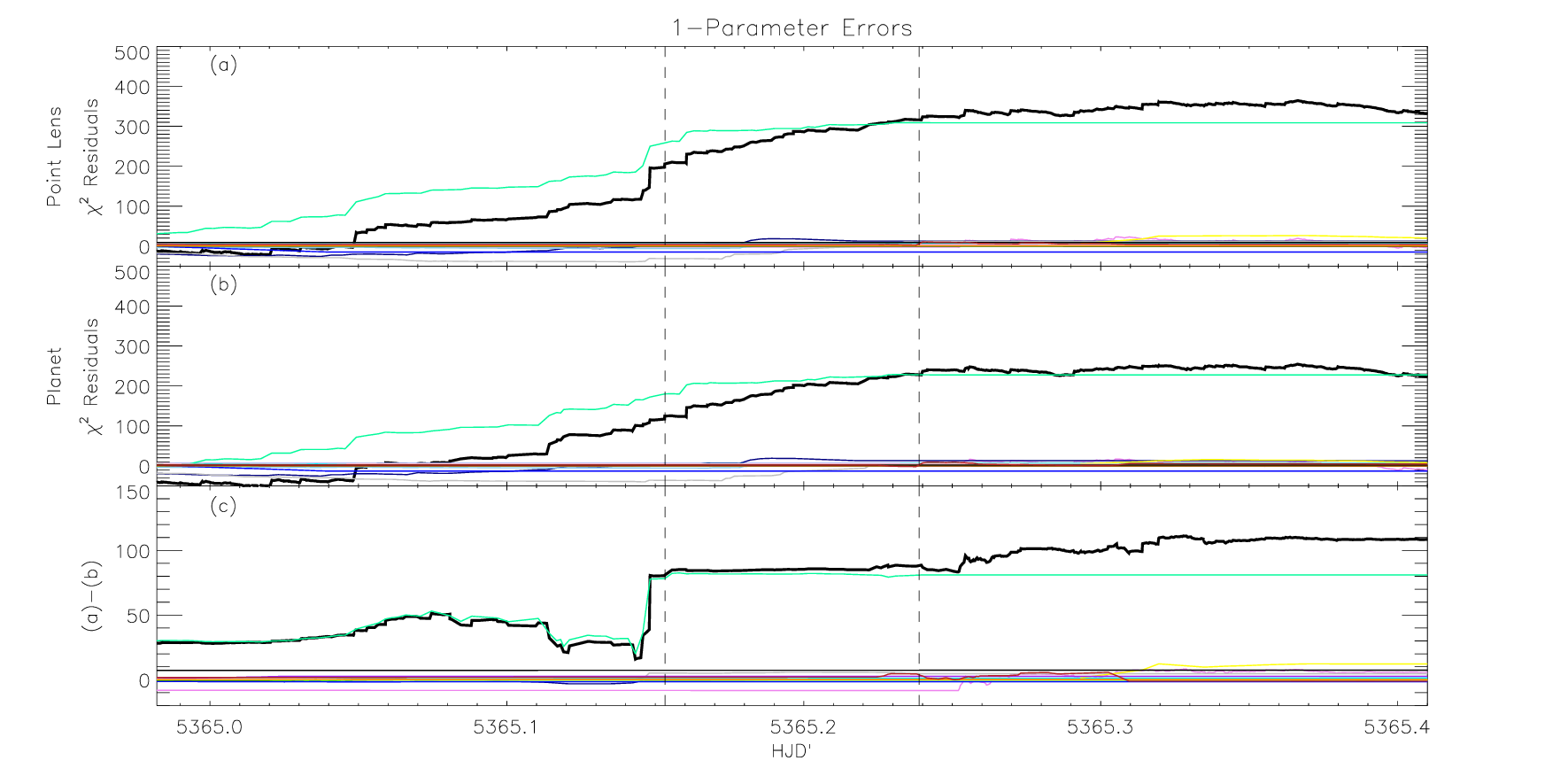}
\includegraphics[width=\textwidth]{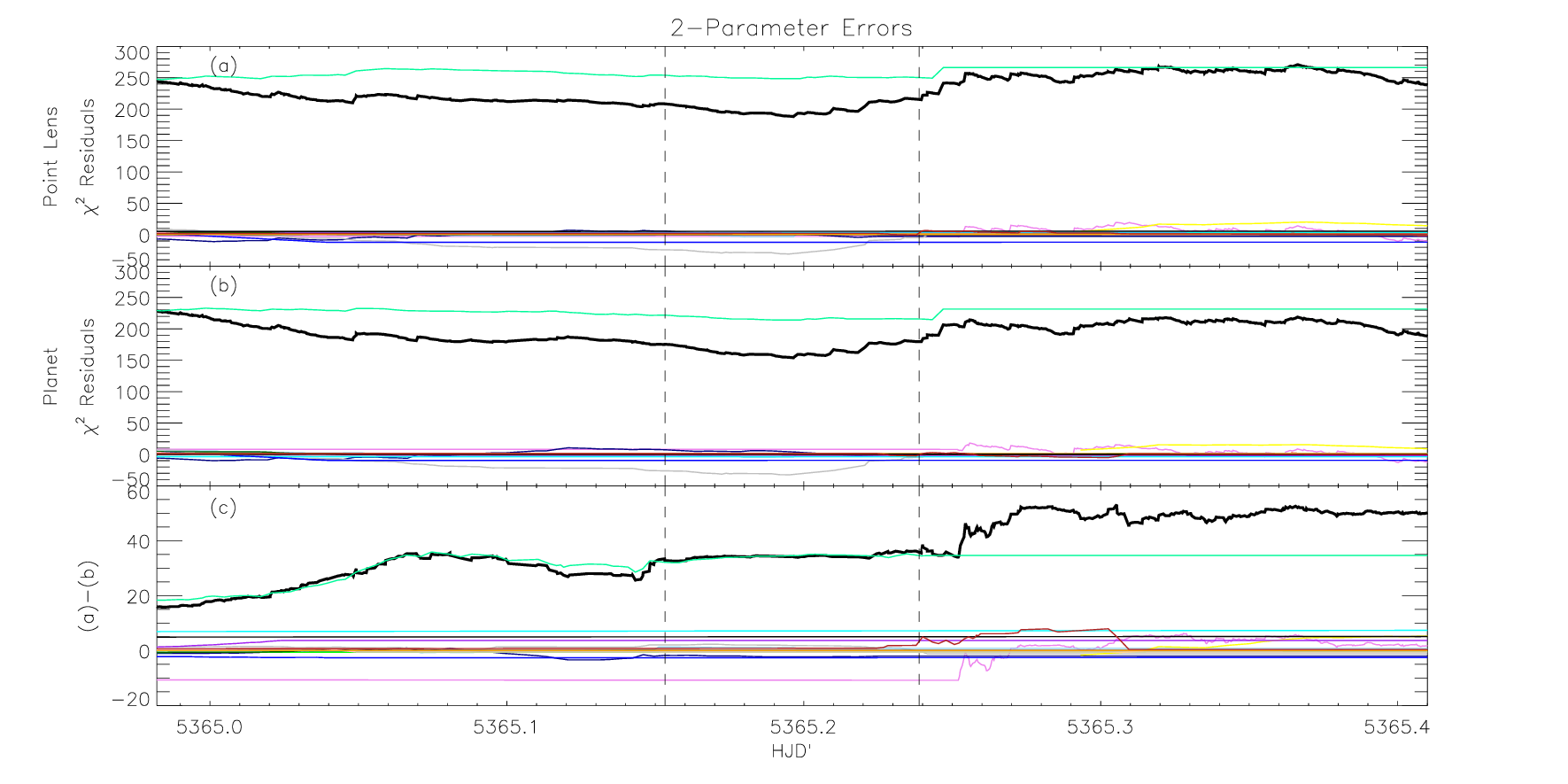}
\caption{The difference between the cumulative $\chi^2$ distribution
and the expected distribution ($\sum_j^N\chi^2_j=N$), i.e. the
$\chi^2$ residuals, for both 1-parameter (top) and 2-parameter
(bottom) errors. The top two panels of each set show the distributions
for the point lens model (a) and the planetary models (b). The bottom
panels (c) of each set show the difference between the top and middle
panels. The distributions for each data set are plotted separately and
are shown over the same time range as the bottom panel of
Fig. \ref{fig:lc}; data sets without points in this time range are not
shown. The colors are as in Fig. \ref{fig:lc}. The thick black line
shows the total distribution for all data. The limb-crossing times are
indicated by the dashed lines. Note the jump in the MOA data
({\it light green}) at the time of the first limb-crossing
(HJD$^{\prime}\sim5361.5$). The signal is much more pronounced for
1-parameter errors than for 2-parameter errors. Note that the vertical
scales in the two sets of panels are different.
\label{fig:cumchi2}}
\end{figure}

To compare the point lens and planetary models, in Figure
\ref{fig:cumchi2}, we plot the ``$\chi^2$ residuals'': the difference
between the cumulative $\chi^2$ distribution and the expected value
$\sum_j^N \chi^2_j=N$, i.e. each point is expected to contribute
$\chi^2_j=1$. For both the point lens and the planet fits, the
distribution rises gradually over the peak of the event. This is
expected since 1-parameter errors do not account for correlated
noise. However, in the $\chi^2$ residuals for the point lens, there is
a jump seen at the time of the first limb-crossing. This jump is even
more pronounced when looking at the difference between the planet and
point lens models. The jump is caused by MOA data at the time of the
limb-crossing that do not fit the point lens well, thereby causing an
excess increase in $\chi^2$. This is exactly the time when we expect
to see planetary signals.

Finally, given the extreme finite source effects in this event, we
might be concerned that uncertainties in the limb-darkening
coefficients due to uncertainties in the source properties could
influence our conclusions. The planetary signal has two components: a
limb-crossing signal and an asymmetry. The limb-darkening could
influence the first signal, but not the second. To check that the
limb-darkening coefficients do not significantly influence our
results, we repeat the point lens fits allowing the limb-darkening
coefficients to be free parameters. In all cases (no parallax,
parallax, 1-parameter or 2-parameter errors), the improvement to the
fit from free limb-darkeing is $\Delta\chi^2\lesssim10$, much smaller
than the planetary signal. Furthermore, the value of $\Gamma_V$
decreases by $>10\%$, which is excluded by the measured source
parameters. Thus, we conclude that our treatment of the limb-darkening
is reasonable.

{\subsection{Reliability of the Planetary Signal}
\label{sec:reliability}}

Although $\Delta\chi^2\sim 140$ appears to be significant, we are
hesitant to claim a detection of a planet. The planetary signal is
more or less equally divided between the jump at the first
limb-crossing and a more gradual rise after the second limb-crossing
(see the third panel of Fig. \ref{fig:cumchi2} showing the difference
between the point lens and planet models). One could argue that the
gradual rise, due to a slight asymmetry in the planet light curve,
could be influenced by large-scale correlations in the data. Comparing
Figures \ref{fig:res} and \ref{fig:cumchi2} shows that most of the
signal at the first limb-crossing comes from only a few points. A
careful examination of the residuals in Figure \ref{fig:res} shows
that while the residuals to the planet model are smaller than for the
point lens model, they are not zero, and the didactic residuals do not
go neatly through the difference between the models as they do for
MOA-2008-BLG-310 \citep{Janczak10}. Hence, the evidence for the planet
is not compelling.

If we repeat the analysis using 2-parameter errors, we find a similar
planetary solution, although the exact values of the parameters are
slightly different\footnote{For 2-parameter errors, model ``A, wide''
appears to be competitive with model ``C''. However, this solution
requires that the source trajectory pass over the planetary caustic at
the exact time to compensate for a night for which the MOA baseline
data are high by slightly more than 1$\sigma$ compared to other nights
at baseline. If the data from this night are removed, the remaining
data predict a different solution with the planetary caustic crossing 18
days earlier. Because this solution is pathological, we do not
consider it further.\label{fn:Awide}}. The total signal from the
planet is significantly degraded for 2-parameter errors, with only
$\Delta\chi^2\sim80$ between the best-fit planet and point lens
models. Table \ref{tab:models2} gives parameters for the complete set
of point lens and planet fits for 2-parameter errors. The residuals
and error bars over peak may be compared to 1-parameter errors in
Figure \ref{fig:res}. 

We also show the $\chi^2$ residuals for 2-parameter errors in the
bottom set of panels in Figure \ref{fig:cumchi2}. They are more or
less flat over the peak, showing that they track the data well in this
region. The offset from zero is caused by systematics elsewhere in the
light curve. The difference plot (bottom-most panel) shows that the
planet fit is still an improvement over the point lens fit, but the
signal from the planet at the first limb-crossing is much weaker. This
is a natural consequence of 2-parameter errors, since the data at the
peak of the event, where the planetary signal is seen, have much
larger renormalized error bars than for 1-parameter
errors\footnote{Note that while the outliers are slightly different
for 1-parameter and 2-parameter errors, no points were rejected in
either case during the first limb-crossing,
$5365.13<t($HJD$^{\prime})<5365.18$, when the main planetary signal is
observed.}.

Regardless of the error renormalization, this planetary signal is
smaller than the $\Delta\chi^2$ of any securely detected
high-magnification microlensing planet. Previously, the smallest
$\Delta\chi^2$ ever reported for a high-magnification event was for
MOA-2008-BLG-310 with $\Delta\chi^2=880$
\citep{Janczak10}. \citet{Yee12} discuss MOA-2011-BLG-293, an event
for which the authors argue the planet could have been detected from
survey data alone with $\Delta\chi^2=500$. However, although the
planet is clearly detectable at this level, it is unclear with what
confidence the authors would have claimed the detection of the planet
in the absence of the additional followup data, which increases
the significance of the detection to $\Delta\chi^2=5400$. At an even
lower level, \citet{Rhie00} find that a planetary companion to the
lens improves the fit to MACHO-98-BLG-35 at $\Delta\chi^2=20$, but
they do not claim a detection. As previously mentioned,
\citet{Gould10} suggest the minimum ``detectable'' planet will have
$350<\Delta\chi^2<700$. However, this threshold has not been rigorously
investigated; the minimum $\Delta\chi^2$ could be smaller.

Because of the tenuous nature of the planetary signal, we do not claim
to detect a planet in this event, but since including a planet in the
fits significantly improves the $\chi^2$, we will refer to this as a
``candidate'' planet.

Finally, it is interesting to note that even though the $\Delta\chi^2$
for the planetary model is too small to be considered detectable, for
both 1-parameter and 2-parameter errors the parameters of the planet
($s$ and $q$) are well defined (see Tables \ref{tab:models1} and
\ref{tab:models2}). Central caustics can be degenerate, especially
when they are much smaller than the source size, so we might expect a
wide range of possible mass ratios in this case since the
limb-crossings are not well-resolved \citep{Han09}. However, perhaps
we should not be surprised that the planet parameters are
well-constrained: both $\Delta\chi^2=80$ and $\Delta\chi^2=140$ are
formally highly significant, which would plausibly lead to reasonable
constraints on the parameters. In this case, because we believe that
the signal could be caused by systematics, by the same token, the
constraints on the parameters may be over-strong. We conjecture that
the limb-crossing signal does not constrain $q$ and that this
constraint actually comes from the asymmetry of the light curve, since
small, central caustics due to planets are asymmetric whereas those
due to binaries are not.

{\section{$\theta_{\rm E}$ and $\mu_{\rm rel}$}
\label{sec:temu}}

Because the source size, $\rho$, is well measured, we can determine
the size of the Einstein ring, $\theta_{\rm E}$, and the relative
proper motion between the source and the lens, $\mu_{\rm rel}$ from the
following relations:
\begin{eqnarray}
\label{eqn:temu}
\theta_{{\rm E}} = \frac{\theta_{\star}}{\rho}&\mathrm{and}&\mu_{\rm
rel} = \frac{\theta_{{\rm E}}}{t_{{\rm E}}}.
\end{eqnarray}
Keeping the limb-darkening parameters fixed, we find the best fit for
the normalized source size to be $\rho = (2.70\pm0.06) \times 10^{-3}$
for the planetary fits; the value is comparable for the point lens
fits. We convert the $(V-I)$ color to $(V-K)$ using \citet{Bessell88}
and obtain the surface brightness by adopting the relation derived by
\citet{Kervella04}. Combining the dereddened $I$ magnitude with this
surface brightness yields the angular source size $\theta_{\star} =
1.03 \pm 0.07\ \mu {\rm as}$. The error on $\theta_{\star}$ combines
the uncertainties from three sources: the uncertainty in flux
($f_{s,I}$), the uncertainty from converting $(V-I)$ color to the
surface brightness, and the uncertainty from the \citet{Nataf12} estimate
of $I_{\rm 0,cl}$. The uncertainty of $f_{s,I}$ is 3\%, which is
obtained directly from the modeling output. We estimate the
uncertainty from the other factors ($Z$) to be 7\%. The fractional
error in $\theta_{\star}$ is given by $[(1/4)(\sigma _{f_{s,I}})^2 +
(\sigma _Z / Z)^2]^{1/2}$ = 7\%, which is also the fractional error
of the proper motion $\mu$ and the Einstein ring radius $\theta_{{\rm
E}}$ \citep{Yee09}. Thus, we find $\theta_{\rm E}=0.38\pm0.03$ mas and
$\mu_{\rm rel}=7.1\pm0.6$ mas yr$^{-1}$.

{\section{The Lens as a Possible Member of NGC 6553}
\label{sec:cluster}}

\begin{figure}
\includegraphics[height=3in]{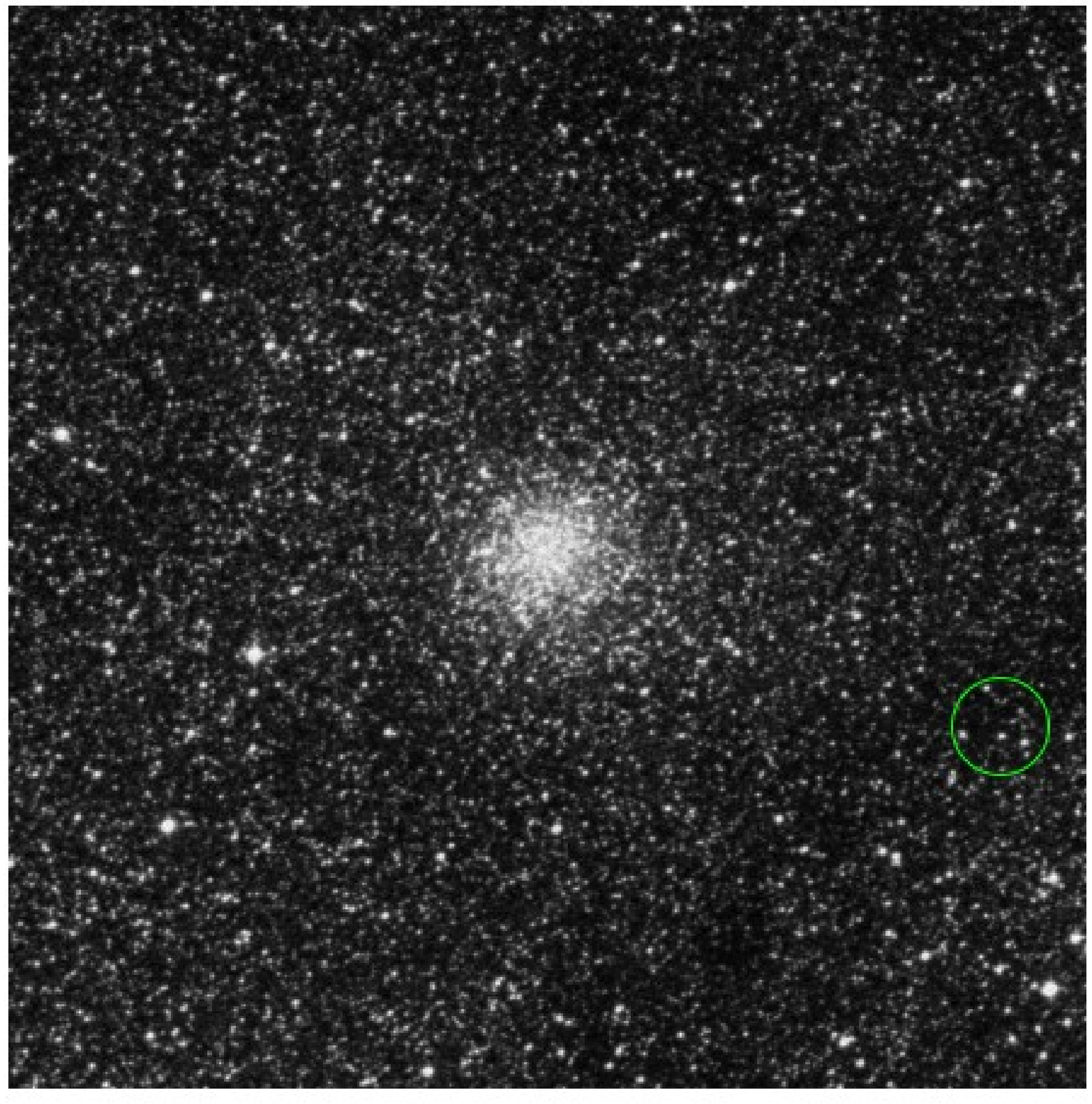}
\includegraphics[height=3in]{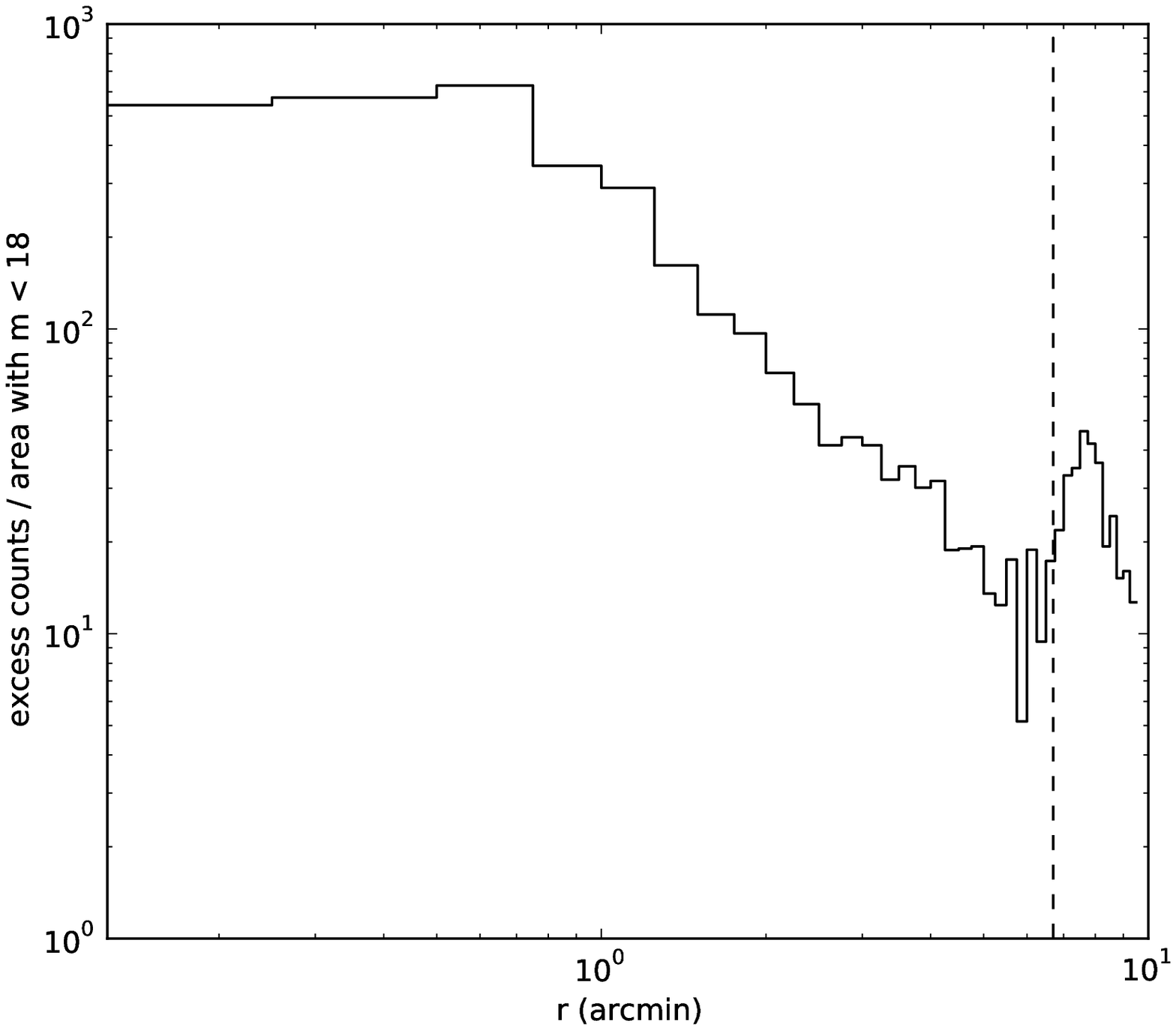}
\caption{DSS image of NGC 6553 (left). The position of the
microlensing event is indicated by the circle. Excess star density
over the background around the globular cluster NGC 6553 (right). The
center of the cluster is placed at $r = 0$. The microlensing event is
at $r = 6.7'$ (dashed line). The dip at $\sim 6'$ is caused by a
quasi-circular dust lane, which may be seen in the image. Even though
the density of stars drops quickly as a function of radius, cluster
members still comprise about 6\% of the stars at $r =
6.7'$. \label{fig:clusterdensity}}
\end{figure}

This microlensing system is close in projection to the globular
cluster NGC 6553. The cluster is at (R.A., decl.) = (18$^{\rm
h}$09$^{\rm m}$17.\hskip-2pt$^{\rm s}$60,
-25$^{\circ}54'31.\hskip-2pt''3$) (J2000.0), with a distance of 6.0
kpc from the Sun and 2.2 kpc from the Galactic Center
\citep{Harris96}. The half-light radius $r_h$ of NGC 6553 is $1.03'$
\citep{Harris96}, which puts the microlensing event 6.5 half-light
radii ($6.7'$) away from the cluster center. By plotting the density
of excess stars over the background, we find that about 6\% of the
stars at this distance are in the cluster (Figure
\ref{fig:clusterdensity}).

Whether or not the lens star is a member of the cluster can be
constrained by calculating the proper motion of the lens star.
\citet{Zoccali01} found a relative proper motion of NGC 6553 with
respect to the Bulge of ${\bdv{\mu}}=(\mu _l,\mu _b) = (5.89, 0.42)$ mas
yr$^{-1}$. The typical motion of Bulge stars is about $100\,\rm
km\,s^{-1}$ corresponding to about $3\,\rm mas\,yr^{-1}$ given an
estimated distance of 7.7 kpc toward the source along this line of sight
\citep[$l=5.1^{\circ}$; ][]{Nataf12}.  Therefore the expected amplitude of the
lens-source relative proper motion if the lens is a cluster member is
$\mu_{\rm rel}=7\pm 3\,\rm mas\,yr^{-1}$, which is consistent with the
measured value in Section (\ref{sec:temu}).

Combining the measurement of the stellar density with the proper
motion information, we find the probability that the lens is a cluster
member is considerably higher than the nominal value based only on
stellar density. First, of order half the stars in the field are
behind the source, whereas the lens must be in front of the
source. Second, the lens-source relative proper motion is consistent
with what would be expected for a cluster lens at much better than
$1\,\sigma$, which is true for only about 2/3 of events seen toward
the Bulge.  Combining these two effects, we estimate a roughly 18\%
probability that the lens is a cluster member.

One way to resolve this membership issue is by measuring the true
proper motion of the lens as was done for a microlensing event in M22
for which the lens was confirmed to be a member of the globular
cluster \citep{Pietrukowicz12}. For this event, we have calculated the
relative proper motion of the lens and the source to be $7.1 \pm 0.6
{\rm\ mas\ yr}^{-1}$. This is consistent with the expected value if
the lens were a member of the cluster. About 10 years after the event,
the separation between the lens and the source star will be large
enough to be measured with $HST$. Based on this followup observation,
one will be able to clarify whether the lens is a member of the
cluster by measuring the vector proper motion. If it is a member of
the cluster, its mass may be estimated from
\begin{equation}
\label{eqn:mass1}
M_{\rm lens} = \frac{\theta_{\rm E}^2}{\kappa \pi_{{\rm rel}}},
\end{equation}    
where $\kappa \equiv 4G/c^2{\rm AU} = 8.14\ {\rm mas} / M_{\sun}$ and
 $\pi_{{\rm rel}}$ is the source-lens relative parallax. In
 addition, the metallicity of the lens could be inferred from the
 metallicity of the globular cluster.

{\section{Discussion}
\label{sec:discuss}}

We have found a candidate planet signal in MOA-2010-BLG-311. The evidence in
support of the planet is
\begin{enumerate}
\item{The planet substantially improves the fit to the data,}
\item{In addition to a general improvement to the light curve, the
planet produces a signal when we most expect it, i.e. the time of the
first limb-crossing,}
\item{The solution has a well-defined mass ratio and projected
separation for the planet (excepting the well known $s\rightarrow
s^{-1}$ degeneracy).}
\end{enumerate}
The magnitude of the signal depends on whether the error bars are
renormalized using 1-parameter ($\Delta\chi^2=140$) or 2-parameter
error factors ($\Delta\chi^2=80$). We conservatively adopt
$\Delta\chi^2=80$ as the magnitude of the signal, but note that if
correlated errors are the dominant source of systematic
uncertainty, $\Delta\chi^2=140$ should be adopted instead (see Section
\ref{sec:errors}). Regardless, this signal is too small to claim
as a secure detection.

Examining the residuals to the light curve and the $\chi^2$ residuals
shows that the planetary signal is dispersed over many points at the
peak of the light curve. It comes from an overall asymmetry near the peak
plus a few points at the limb-crossing. Because the signal is the sum
of multiple cases of low-amplitude deviations, it is plausible that the
microlensing model could be fitting systematics in the data, which is
why the planet signal is not reliable. 

Combined with other studies, this event suggests that central-caustic
(high-magnification) events and planetary-caustic events require
different detection thresholds. The detection threshold suggested in
\citet{Gould10} of $350<\Delta\chi^2<700$ was made in the context of
high-magnification events, and our experience so far is consistent
with this. A planet was clearly detectable in MOA-2008-BLG-310 with
$\Delta\chi^2=880$ \citep{Janczak10}. However, \citet{Yee12} are
uncertain if a planet would be claimed with $\Delta\chi^2=500$ for
MOA-2011-BLG-293. Here, $\Delta\chi^2=80$ is definitely insufficient
to detect a planet. Hence, the detection threshold for planets in
high-magnification events is around or just below
$\Delta\chi^2=500$. In contrast, the planetary caustic crossing in
OGLE-2005-BLG-390 produced a clear signal of $\Delta\chi^2=532$
\citep{Beaulieu06}, and the planet would most likely be detectable if
the error bars were 50\% larger ($\Delta\chi^2\sim200$) and might even
be considered reliable if the error bars were twice as large.

We suspect the reason for the different detection thresholds is that
the information about the microlens parameters and the planetary
parameters comes from different parts of the light curve. For
planetary-caustic events, the planet signal is a perturbation
separated from the main peak. Thus, the microlens parameters can be
determined from the peak data independently from the planetary
parameters, which are measured from the separate, planetary
perturbation. In contrast, for high-magnification events, the
planetary perturbation occurs at the peak of the event, so the
microlens and planetary parameters must be determined from the same
data.

The detection threshold for planetary-caustic events will have to be
investigated in more detail. If it is truly lower than for
high-magnification events, this is good news for second generation
microlensing surveys since that is how most planets will be found in
such surveys. At the same time, it points to the continued need for
followup data of high-magnification events since these seem to have a
higher threshold for detection, requiring more data to confidently
claim a planet. This is an important consideration because
high-magnification events can yield much more detailed information
about the planets.

Additionally, \citet{HanKim09} show that the magnitude of the
planetary signal should decrease as the ratio between the caustic size
and the source size ($w/\rho$) decreases for the same photometric precision\footnote{\citet{Chung05} give
equations for calculating the caustic size, $w$ \citep[see also
][]{Dong09_400}.}. There are
several cases of events for which $w/\rho\lesssim 1$: in this case, we
have $w/\rho=0.12$ and $\Delta\chi^2=140$; the brown dwarf in
MOA-2009-BLG-411L has $w/\rho=0.3$ and $\Delta\chi^2=580$
\citep{Bachelet12b}; MOA-2007-BLG-400 has $w/\rho=0.4$ and
$\Delta\chi^2=1070$ \citep{Dong09_400}; and in the case of
MOA-2008-BLG-310, the value is $w/\rho=1.1$ with $\Delta\chi^2=880$
\citep{Janczak10}. This sequence is imperfect, but the photometry in
the four cases is far from uniform, and it seems that in general the
trend suggested by \citet{HanKim09} holds in practice.

Finally, we note that a large fraction of the planet signal comes from
the MOA data, but in the original, real-time MOA data, this signal
would not have been detectable since the data were corrupted over the
peak of the event. It is only after the data quality was improved by
rereductions, which in turn were undertaken only because the event
became the subject of a paper, that we recovered the planet candidate. This
points to the importance of rereductions of the data when searching
for small signals. In the current system of analyzing only the events
with the most obvious signals, this is not much of a concern. However,
if current or future microlensing surveys are systematically analyzed
to find signals of all sizes, this will become important.

{\large\bf Acknowledgments} 

\acknowledgements{The MOA collaboration acknowledges the support of
grants JSPS20340052 and JSPS22403003. The OGLE project has received
funding from the European Research Council under the European
Community's Seventh Framework Programme (FP7/2007-2013) / ERC grant
agreement no. 246678 to A. Udalski. Work by J.C. Yee was supported by
a National Science Foundation Graduate Research Fellowship under Grant
No. 2009068160. A. Gould and J.C. Yee acknowledge support from NSF
AST-1103471. B.S. Gaudi, A. Gould, L.-W. Hung, and R.W. Pogge
acknowledge support from NASA grant NNX12AB99G. A. Gould and D. Maoz
acknowledge support by a grant from the US Israel Binational Science
Foundation. Work by C. Han was supported by Creative Research
Initiative Program (2009-0081561) of National Research Foundation of
Korea. Work by S. Dong was performed under contract with the
California Institute of Technology (Caltech) funded by NASA through
the Sagan Fellowship Program. Mt. Canopus Observatory is supported by
Dr. David Warren. T.C. Hinse gratefully acknowledges financial support
from the Korea Research Council for Fundamental Science and Technology
(KRCF) through the Young Research Scientist Fellowship
Program. T.C.Hinse and C.-U. Lee acknowledge financial support from
KASI (Korea Astronomy and Space Science Institute) grant number
2012-1-410-02. B. Shappee and J. van Saders are also supported by
National Science Foundation Graduate Research Fellowships. K. Alsubai,
D.M. Bramich, M. Dominik, K. Horne, M. Hundertmark, C. Liebig,
C. Snodgrass, R.A. Street and Y. Tsapras would like to thank the Qatar
Foundation for support from QNRF grant NPRP-09-476-1-078. The MiNDSTEp
monitoring campaign is powered by ARTEMiS (Automated Terrestrial
Exoplanet Microlensing Search) [Dominik et al. 2008, AN 329,
248]. M. Hundertmark acknowledges support by the German Research
Foundation (DFG). D. Ricci (boursier FRIA) and J. Surdej acknowledge
support from the Communaut/'e fran\c{c}aise Belgique Actions de
recherche concert\'ees -- Acad/'emie universitaire
Wallonie-Europe. C. Snodgrass received funding from the European Union
Seventh Framework Programme (FPT/2007-2013) under grant agreement
268421. This work is based in part on data collected by MiNDSTEp with
the Danish 1.54 m telescope at the ESO La Silla Observatory. The
Danish 1.54 m telescope is operated based on a grant from the Danish
Natural Science Foundation (FNU).

The Digitized Sky Surveys were produced at the Space Telescope Science
Institute under U.S. Government grant NAG W-2166. The images of these
surveys are based on photographic data obtained using the Oschin
Schmidt Telescope on Palomar Mountain and the UK Schmidt
Telescope. The plates were processed into the present compressed
digital form with the permission of these institutions. The National
Geographic Society - Palomar Observatory Sky Atlas (POSS-I) was made
by the California Institute of Technology with grants from the
National Geographic Society. The Second Palomar Observatory Sky Survey
(POSS-II) was made by the California Institute of Technology with
funds from the National Science Foundation, the National Geographic
Society, the Sloan Foundation, the Samuel Oschin Foundation, and the
Eastman Kodak Corporation. The Oschin Schmidt Telescope is operated by
the California Institute of Technology and Palomar Observatory. The UK
Schmidt Telescope was operated by the Royal Observatory Edinburgh,
with funding from the UK Science and Engineering Research Council
(later the UK Particle Physics and Astronomy Research Council), until
1988 June, and thereafter by the Anglo-Australian Observatory. The
blue plates of the southern Sky Atlas and its Equatorial Extension
(together known as the SERC-J), as well as the Equatorial Red (ER),
and the Second Epoch [red] Survey (SES) were all taken with the UK
Schmidt.  }



\begin{deluxetable}{ll|lr|llr|}

\tablecaption{Data Properties for Two Methods of Error
Renormalization. The observatories are listed in order of longitude
starting with the most Eastward. If data were taken in more than one
filter at a given site, different filters are given on successive
lines. The error renormalization coefficients and method for removing
outliers are described in Section \ref{sec:errors}.\label{tab:data}}

\tablehead{\colhead{}&\colhead{}&\multicolumn{2}{c}{1-parameter Errors}&\multicolumn{3}{c}{2-parameter Errors}\\
\colhead{Observatory}&\colhead{Filter}&\colhead{$k$}&\colhead{$N_{\rm data}$}&\colhead{$k$}&\colhead{$e_{\rm min}$}&\colhead{$N_{\rm data}$}\\
}
\startdata
Mt. Bigelow  &I            &  1.63&   44& 1.63& 0.0   &  44\\
Molehill     &Unfiltered   &  0.72&   69& 0.72& 0.0   &  69\\
Kumeu        &Wratten \#12 &  1.19&  188& 1.18& 0.0   & 188\\
Farm Cove    &Unfiltered   &  1.27&   52& 1.26& 0.0   &  52\\
Auckland     &Wratten \#12 &  0.98&   84& 1.00& 0.0   &  84\\
Vintage Lane &Unfiltered   &  4.87&  112& 3.05& 0.004 & 112\\
B\&C         &I            &  4.07&  132& 1.01& 0.025 & 136\\
             &V            &  1.15&   53& 0.66& 0.03  &  55\\
MOA          &MOA-Red      &  1.68& 4452& 1.55& 0.003 &4434\\
Canopus      &I            &  3.02&   28& 2.87& 0.0   &  29\\
Wise         &Unfiltered   &  0.52&   70& 0.53& 0.0   &  70\\
Bronberg     &Unfiltered   &  1.26&  727& 1.27& 0.0   & 727\\
SAAO         &I            &  2.60&  128& 2.21& 0.0015& 127\\
Liverpool    &SDSS-i       &  1.85&  120& 1.04& 0.007 & 119\\
La Silla     &I            & 10.02&  169& 3.79& 0.004 & 174\\
CTIO         &I            &  1.33&   22& 1.34& 0.0   &  22\\
             &V            &  0.50&    3& 0.50& 0.0   &   3\\
          &H\tablenotemark{a}&\nodata&74& \nodata&\nodata&74\\
OGLE         &I            &  1.36&  429& 1.32& 0.008 & 429\\
\enddata

\tablenotetext{a}{These data were not used in the modeling. They were
only used to determine the source color (see Sec. \ref{sec:cmd}).}

\end{deluxetable}

\begin{deluxetable}{lrlllllllrr}

\tablecaption{Fits with 1-Parameter Errors. The $\Delta\chi^2$ is
given relative to the $\chi^2$ of the best-fit planetary model with
$s<1$ and without parallax ($\chi^2=6637.96$), i.e. model ``C'';
positive numbers indicate a worse fit and negative numbers indicate an
improvement relative to this model. The point lens models are given
first, followed by the planetary models; ``A'', ``B'', and ``C''
denote the three planetary models with distinct values of $\alpha$
corresponding to the three $\chi^2$ minima. \label{tab:models1}}

\rotate
\tabletypesize{\scriptsize}
\setlength{\tabcolsep}{0.04in}

\tablehead{\colhead{Model}&\colhead{$\Delta\chi^2$}&\colhead{$t_0-5365.$}&
\colhead{$u_0$}&\colhead{$t_{\rm E}$}&\colhead{$\rho$}&\colhead{$\alpha$}&
\colhead{$\log s$}&\colhead{$\log q$}&\colhead{$\pi_{E,N}$}&
\colhead{$\pi_{E,E}$}\\

\colhead{}&\colhead{}&\colhead{(HJD$^{\prime}$)}&\colhead{}&\colhead{(days)}&\colhead{}&\colhead{($^{\circ}$)}&\colhead{}&\colhead{}}
\startdata
          Point Lens&  136.44&0.19615(4) & \phs0.00152(3) & 20.34(42) &0.00260(5) &    \nodata&    \nodata&    \nodata&    \nodata&  \nodata\\
        PL, parallax&   69.61& 0.1978(2) & \phs0.00167(4) & 19.37(41) &0.00275(6) &    \nodata&    \nodata&    \nodata&  3.16(41) &-1.34(20)\\
PL, parallax, $-u_0$&  123.89&0.19613(9) & \phn-0.00158(4) & 20.23(42) &0.00262(5) &    \nodata&    \nodata&    \nodata& -1.07(36) &-0.98(27)\\
\hline
\hline
                   A&   35.26&0.19615(4) & \phs0.00167(3) & 19.04(34) &0.00279(5) &\phs347.7(6) &\phn-0.12(1) &-4.46(8) &    \nodata&  \nodata\\
         A, parallax&   -7.02& 0.1976(2) & \phs0.00172(4) & 18.92(40) &0.00283(6) &\phs347.4(3) &\phn-0.08(1) &-4.9(1) &  2.82(44) &-1.29(22)\\
 A, parallax, $-u_0$&   -3.98&0.19626(9) &\phn-0.00184(5) & 18.51(38) &0.00289(6) &\phn-346.(1) &\phn-0.17(2) &  -4.17(8) & -2.11(46) &-2.43(40)\\
             A, wide&   33.07&0.19615(4) & \phs0.00167(4) & 19.06(39) &0.00279(6) &\phs348.1(6) &  \phs0.12(1) &  -4.44(8) &    \nodata&  \nodata\\
\hline
                   B&   49.06&0.19614(4) & \phs0.00159(4) & 19.71(43) &0.00269(6) &\phs118(1) &\phn-0.43(4) &-3.5(1) &    \nodata&  \nodata\\
         B, parallax&   21.45& 0.1973(2) & \phs0.00166(4) & 19.40(41) &0.00275(6) &\phs119(2) &\phn-0.40(5) &-3.6(2) &  2.21(45) &-1.05(21)\\
 B, parallax, $-u_0$&   31.33&0.19617(9) &\phn-0.00169(4) & 19.46(41) &0.00274(6) &\phn-115(1)&\phn-0.40(3) &-3.5(1) & -1.45(41) &-1.50(35)\\
             B, wide&   49.11&0.19615(4) & \phs0.00159(3) & 19.76(38) &0.00268(5) &\phs118(1) &\phs0.43(4) &-3.5(1) &    \nodata&  \nodata\\
\hline
                   C&    0.00&0.19613(4) & \phs0.00159(3) & 19.68(41) &0.00270(6) &\phs236.4(7)&\phn-0.26(4) &-3.7(1) &    \nodata&  \nodata\\
         C, parallax&  -10.65& 0.1968(3) & \phs0.00164(4) & 19.34(39) &0.00275(6) &\phs235(1)  &\phn-0.4(1) &-3.4(3) &  0.89(51) &-0.78(24)\\
 C, parallax, $-u_0$&  -12.63&0.19630(9) &\phn-0.00171(4) & 19.20(38) &0.00278(6) &\phn-232(1) &\phn-0.51(8) &-3.0(2) & -1.11(46) &-1.55(41)\\
             C, wide&    0.00&0.19613(4) & \phs0.00159(3) & 19.73(39) &0.00269(5) &\phs236.4(7)&\phs 0.26(4) &-3.7(1) &    \nodata&  \nodata\\

\enddata
\end{deluxetable}

\begin{deluxetable}{lrlllllllrr}
\tablecaption{Fits with 2-Parameter Errors. The $\Delta\chi^2$ is
given relative to the $\chi^2$ of the best-fit planetary model with
$s<1$ and without parallax ($\chi^2=6751.93$), i.e. model ``C''; positive
numbers indicate a worse fit and negative numbers indicate an
improvement relative to this model. The point lens models are given
first, followed by the planetary models; ``A'', ``B'', and ``C''
denote the three planetary models with distinct values of $\alpha$
corresponding to the three $\chi^2$ minima. \label{tab:models2}}

\rotate
\tabletypesize{\scriptsize}
\setlength{\tabcolsep}{0.04in}

\tablehead{\colhead{Model}&\colhead{$\Delta\chi^2$}&\colhead{$t_0-5365.$}&\colhead{$u_0$}&\colhead{$t_{\rm
E}$}&\colhead{$\rho$}&\colhead{$\alpha$}&\colhead{$\log
s$}&\colhead{$\log q$}&\colhead{$\pi_{E,N}$}&\colhead{$\pi_{E,E}$}\\

\colhead{}&\colhead{}&\colhead{(HJD$^{\prime}$)}&\colhead{}&\colhead{(days)}&\colhead{}&\colhead{($^{\circ}$)}&\colhead{}&\colhead{}}
\startdata
          Point Lens&   81.12&0.19618(5) & \phs0.00152(3) & 20.51(44) &0.00259(6) &      \nodata&       \nodata&    \nodata&    \nodata&  \nodata\\
        PL, parallax&   25.43& 0.1978(2) & \phs0.00164(4) & 19.80(43) &0.00270(6) &      \nodata&       \nodata&    \nodata&  3.20(43) &-1.14(21)\\
PL, parallax, $-u_0$&   72.23& 0.1961(1) &\phn-0.00160(4) & 20.23(44) &0.00263(6) &      \nodata&       \nodata&    \nodata& -1.26(42) &-0.96(32)\\
\hline
\hline
                   A&    6.37&0.19611(5) & \phs0.00160(4) & 19.82(43) &0.00269(6) &  \phs354(2) & \phn-0.49(5) &   -3.1(1) &    \nodata&  \nodata\\
         A, parallax&  -12.98& 0.1972(3) & \phs0.00168(5) & 19.25(46) &0.00278(7) &  \phs349(3) &  \phn-0.3(2) &   -3.7(6) &  1.89(70) &-1.16(25)\\
 A, parallax, $-u_0$&   -8.07& 0.1963(1) &\phn-0.00172(5) & 19.27(42) &0.00278(6) & \phn-348(2) & \phn-0.46(7) &   -3.1(2) & -0.98(52) &-1.59(48)\\
             A, wide\tablenotemark{1}&   -1.59&0.19614(5) & \phs0.00157(3) & 20.05(37) &0.00266(5) &  \phs356(3) & \phs0.539(4) &  -3.01(3) &    \nodata&  \nodata\\
\hline
                   B&   13.25&0.19613(5) & \phs0.00160(3) & 19.79(41) &0.00270(6) &  \phs114(2) & \phn-0.51(6) &   -3.1(2) &    \nodata&  \nodata\\
         B, parallax&   -6.42& 0.1972(2) & \phs0.00165(4) & 19.54(41) &0.00274(6) &  \phs112(3) &  \phn-0.5(1) &   -3.1(3) &  1.97(48) &-1.02(23)\\
 B, parallax, $-u_0$&    4.22& 0.1962(1) &\phn-0.00167(5) & 19.54(41) &0.00274(6) & \phn-109(3) & \phn-0.50(6) &   -3.1(2) & -0.99(47) &-1.24(43)\\
             B, wide&   13.10&0.19613(5) & \phs0.00159(4) & 19.82(42) &0.00269(6) &  \phs113(2) &  \phs0.51(6) &   -3.1(2) &    \nodata&  \nodata\\
\hline
                   C&    0.00&0.19614(5) & \phs0.00158(3) & 19.94(40) &0.00268(5) &  \phs234(1) & \phn-0.40(7) &   -3.3(2) &    \nodata&  \nodata\\
         C, parallax&  -14.90& 0.1970(3) & \phs0.00164(4) & 19.51(41) &0.00274(6) &  \phs231(2) & \phn-0.56(8) &   -2.9(2) &  1.26(54) &-0.96(24)\\
 C, parallax, $-u_0$&  -10.15& 0.1963(1) &\phn-0.00166(5) & 19.49(42) &0.00274(6) & \phn-230(2) & \phn-0.51(7) &   -2.9(2) & -0.44(51) &-1.04(46)\\
             C, wide&   -0.02&0.19613(5) & \phs0.00158(3) & 19.97(42) &0.00267(6) &  \phs234(1) &  \phs0.40(7) &   -3.3(2) &    \nodata&  \nodata\\

\enddata

\tablenotetext{1}{This solution
and its parameters should be treated with caution, since it
corresponds to a pathological geometry. See footnote \ref{fn:Awide}.}

\end{deluxetable}
\end{document}